# From Informal Addresses to Reliable Places: Participatory Data Governance of Civic Addressing in Puerto Rico[1]

**Juan A. Padilla**
Computer Science and Engineering Department
University of Puerto Rico
Mayagüez, PR, USA
juan.padilla11@upr.edu

## ABSTRACT

This paper examines civic addressing as a problem of participatory data governance. Drawing on a project developed through the U.S. Census Bureau's The Opportunity Project with engagement from FEMA, we describe the use of actionable geolocations to support services where formal addresses are absent. We introduce Reliable Places as transitional governance artifacts through which place reliability emerges via use, enabling services while supporting pathways toward formal civic address assignment.

## INTRODUCTION: CIVIC ADDRESSING AS A GOVERNANCE CHALLENGE

Civic addressing is critical infrastructure for emergency response, service delivery, and public administration. Yet it is often treated as a purely administrative task: streets are named, numbers are assigned, and addresses are assumed to function once recorded in official registries. In practice, this assumption breaks down in contexts where addressing is informal, incomplete, or historically uneven. In Puerto Rico, it is estimated that approximately 40%—roughly 600,000 households—lack a formal street name or house number [1]. The consequences include limited access to emergency services, delivery, and public assistance. Prior work in HCI and STS has shown that infrastructures often become visible only through breakdown, revealing embedded assumptions about access and use rather than purely technical failure [2]. Civic addressing systems similarly function as classificatory infrastructures whose consequences extend beyond navigation to questions of recognition and legitimacy [3].

Rather than framing this condition as missing or low-quality data, we approach it as a governance gap concerning who has the capacity and legitimacy to define place and stabilize representations over time.

## PROJECT CONTEXT: ACTIONABLE GEOLOCATIONS FOR CIVIC ADDRESSING

This paper draws on a project developed in collaboration with the U.S. Census Bureau's The Opportunity Project (TOP), an open innovation challenge led by Census and HUD, with engagement from FEMA and other federal agencies, aimed at enabling service delivery in the absence of formal civic addresses [4]. The immediate objective was not to replace official addressing systems, but to support actionable, community-grounded representations of place while longer-term formalization efforts remain ongoing.

The project leveraged FEMA's USA Structures dataset as a structural baseline, providing a representation of physical dwellings independent of street names or house numbers. These structures were treated as provisional anchors rather than authoritative addresses. Through collaboration with a community organization, residents' actionable identifiers—such as phone numbers and email addresses—were associated with specific structures during on-the-ground engagement. The resulting pairings formed actionable geolocations: place representations capable of supporting communication, coordination, and service delivery despite the absence of formal addresses.

### Institutional Engagement and Deployment Context

The system and associated workflow were developed and shared with the U.S. Census Bureau, FEMA, and other federal stakeholders as part of exploratory efforts to address civic addressing challenges in Puerto Rico. As with many cross-agency initiatives, implementation depends on timing, coordination, and institutional prioritization. The work contributes design insights and operational pathways that can inform future deployments under evolving administrative contexts.

## PARTICIPATION THROUGH USE, NOT CONSULTATION

A central characteristic of this work is how participation occurred in practice. Rather than relying on workshops or consultative feedback alone, community participation was embedded directly into the operational use of the system. Community organizations conduct visits within neighborhoods, associate residents with structures through real interactions tied to anticipated services and communications. This mode of engagement aligns with long-standing arguments in participatory design that participation need not be limited to formal consultation, but can instead emerge through sustained involvement in shaping and maintaining shared systems in practice [5].

[1] This paper is a preprint of a workshop paper accepted at the CHI 2026 Workshop on Participatory Data Governance at the CHI 2026 Conference on Human Factors in Computing Systems.

Participation thus manifests not as opinion-giving but as co-production of operational place data. Reliability emerges gradually, through interaction, rather than being assumed at the moment of data entry.

## TOOLS FOR PARTICIPATORY GOVERNANCE OF PLACE

The system included a map-based interface designed to support collective action around place. Organizers could select geographic areas directly on the map, automatically retrieving all actionable identifiers associated with the enclosed structures. Beyond visualization, the tool enabled the creation of surveys, messages, polls, and votes targeted to specific sets of structures or neighborhoods.

These affordances functioned as mechanisms of participatory governance rather than engagement features alone. Communities could be contacted collectively, respond to shared prompts, express preferences, and surface disagreement tied explicitly to place. In doing so, the system supports ongoing negotiation over how places are represented, which associations are valid, and how collective decisions—such as naming—are made. Governance occurs through the ability to act together on shared place representations. In this sense, the system functions as an instance of *infrastructuring*, where tools for communication and coordination actively shaped how publics formed around shared representations of place, rather than merely supporting preexisting governance structures [6].

## NAMING AS GOVERNANCE: AUTHORITY, IDENTITY, AND LEGITIMACY

An important dimension of this work concerned naming authority. Communities frequently expressed preferences for naming streets or areas based on locally meaningful references, including respected residents, historical events, or iconic natural features. These naming practices were not merely symbolic; they encoded collective memory, identity, and legitimacy. This perspective resonates with HCI accounts of place as relational and enacted through practice, rather than a fixed container defined solely by formal spatial representations [7].

Treating naming as a purely administrative step risks erasing these meanings and undermining community ownership of place data. In this context, naming became a site of participatory data governance, where local knowledge and institutional requirements had to be reconciled rather than hierarchically imposed. The negotiation of names revealed how place identity is socially produced and why legitimacy matters for the durability of civic addressing systems.

## FROM ACTIONABLE GEOLOCATIONS TO RELIABLE PLACES

Across these interactions, a key observation emerged: some places consistently supported successful communication and service interactions, while others did not. Usability was not determined by formal recognition but by reliability in practice. This insight motivated the concept of Reliable Places—places whose trustworthiness is established through repeated, successful use rather than authoritative designation alone. Consistent with research on situated action, reliability here was not the outcome of predefined plans or authoritative designation, but of repeated, context-sensitive interactions that stabilized expectations over time [8]. This interactional view parallels HCI accounts of trust as something produced and maintained through ongoing engagement rather than declared at design time [9].

In this framing, Reliable Places function as transitional governance artifacts. They enable services and participation in the present while accumulating evidence needed for eventual formal civic address assignment. Importantly, they are not alternatives to civic addresses but pathways toward them, allowing institutional processes to build upon stabilized, community-grounded practice.

## FORMAL CIVIC ADDRESSING AS A PARTICIPATORY OUTCOME

The long-term goal of this work is to support the assignment of formal civic addresses to all households on the island. However, this project suggests that durable formalization depends on operational grounding. Top-down address assignment without evidence of usability risks producing formally correct but practically unreliable data.

By engaging communities through actionable geolocations and participatory interaction, the project establishes conditions under which formal civic addresses can be assigned, maintained, and trusted. Formalization, in this sense, becomes an outcome of participatory governance rather than a prerequisite for it.

## IMPLICATIONS FOR PARTICIPATORY DATA GOVERNANCE

This case highlights several implications for participatory data governance in practice. First, participation can occur through use, validation, and collective interaction—not only through formal consultation. Second, authority over place is negotiated and gradual, emerging through stabilized practice rather than instant designation. Finally, systems must accommodate evolving representations of place that balance cultural meaning with operational reliability. Together, these observations suggest that participatory data governance must accommodate evolving, negotiated forms of classification and trust that emerge through use, particularly in contexts where formal infrastructures remain incomplete or uneven.

## QUESTIONS FOR WORKSHOP DISCUSSION

• When does interaction constitute governance in data systems?

• How should community-defined naming coexist with institutional standards?

• What forms of evidence should support the transition from provisional to formal place representations?

• How can participatory governance of place scale without erasing local specificity?